\documentclass[aps,prx,twocolumn,showpacs,superscriptaddress]{revtex4-2}

\usepackage[dvipdfmx]{graphicx}
\usepackage{siunitx}
\usepackage[version=4]{mhchem}

\begin{document}

\title{Accelerating Optimal Elemental Configuration Search in Crystal using Ising Machine}

\author{Kazuhide Ichikawa}
\affiliation{Technology Division, Panasonic Holdings Corporation, Osaka 571-8508, Japan}

\author{Satoru Ohuchi}
\affiliation{Technology Division, Panasonic Holdings Corporation, Osaka 571-8508, Japan}

\author{Koki Ueno}
\affiliation{Technology Division, Panasonic Holdings Corporation, Osaka 571-8508, Japan}

\author{Tomoyasu Yokoyama}
\affiliation{Technology Division, Panasonic Holdings Corporation, Osaka 571-8508, Japan}

\date{May 31, 2023}

\begin{abstract}
This research demonstrates that Ising machines can effectively solve optimal elemental configuration searches in crystals, with Au-Cu alloys serving as an example. 
The energy function is derived using the cluster expansion method in the form of a QUBO function, enabling efficient problem-solving via Ising machines. 
We have successfully obtained reasonable solutions for crystal structures consisting of over 10,000 atoms. 
Notably, we have also obtained plausible solutions for optimization problems with constrained solutions, such as situations where the composition ratio of atomic species is predetermined. 
These findings suggest that Ising machines can be valuable tools for addressing materials science challenges.
\end{abstract}

\maketitle

\section{Introduction}
In recent times, the advancement of materials has increasingly relied on calculations to support the physical properties of synthesized materials and to forecast the properties of undiscovered materials. Specifically, first-principles calculations are extensively employed to ascertain various physical properties of materials once their structure (the arrangement of atoms that constitute the material) is known. Nonetheless, because the structure of a material that can be synthesized is energetically stable, it is essential to understand the stable crystal structure of the material to predict its physical properties through calculations. In many instances, the stable structure of a material can be closely approximated by the structure possessing the least energy. However, in a multinary crystal system, which comprises a collection of atoms from more than one type, it is frequently the case that only the positions where any of the atoms can be situated (subsequently referred to as sites or site positions) are known. As a result, determining which atoms (or vacancies) should be positioned at which sites in the crystal to establish a stable structure often remains uncertain.

For instance, oxides of transition metal elements are frequently utilized as cathode materials in lithium-ion secondary batteries. Performance, including potential and capacity, can be enhanced by incorporating multiple species of transition metal elements, with the extent of improvement being dependent on the proportion of these species. To computationally determine the ideal ratio for optimal performance, it is initially essential to identify the placement of transition metals at specific crystal sites to ensure structural stability. Consequently, a method for determining the stable structure is vital for the progression of material development. Furthermore, the significance of the method for establishing site assignments is growing, given the recent introduction of multinary materials for a diverse range of applications, encompassing anode materials, solid electrolyte materials, high-temperature structural materials, biological materials, fusion reactor structural materials, electrolytic capacitor materials, and catalyst materials.

A traditional approach to identifying stable crystal structures involves determining the optimal arrangement of individual atoms based on energy considerations. This procedure necessitates calculating the energy for every possible configuration within the crystal structure and then choosing the combination that results in the lowest energy. In this paper, we call this problem the elemental configuration optimization problem.

Two challenges arise when addressing the elemental configuration optimization problem: the high computational cost of first-principles calculations for each energy, and the exponential growth in the number of configuration combinations relative to the number of atoms. In traditional materials computational science, the ``cluster expansion method" \cite{SANCHEZ1984334, de1994solid, ceder2000first} has been employed as a means of mitigating the first issue by reducing the cost of energy calculations. To tackle the second challenge, simulated annealing has been utilized \cite{ozolicnvs1998cu, seko2006prediction,chang2019clease}. For instance, these techniques have recently been applied to predict the La/Li/vacancy configuration of lithium-ion conducting materials \cite{yang2020arrangement}.

This research paper emphasizes the potential acceleration of the second challenge through the adoption of Ising machines. These specialized computers are designed for solving combinatorial optimization problems and have emerged in recent years \cite{johnson2011quantum,aramon2019physics, nakayama2021description, yamamoto20211, takemoto20214, goto2019combinatorial, yamamoto20211,tanaka2020theory}.
In Section~\ref{sec:methods}, the cluster expansion method is outlined, and it is described that the energy function can be obtained in a format that can be optimized by Ising machines. In Section~\ref{sec:results}, we present the calculation results using the Cu-Au alloy as an example and demonstrate that stable structures are successfully identified for each composition ratio. Finally, in the concluding section, we provide a summary and discuss future prospects.

\section{Methods}  \label{sec:methods}

In this section, we provide an outline of the cluster expansion method employed in materials computational science. For the sake of simplicity, we will only discuss how, in the cluster expansion method, the energy of the crystal structure is presented in a format that is well-suited for optimization by Ising machines. For more comprehensive information, readers are referred to the cited references \cite{SANCHEZ1984334, de1994solid, ceder2000first, wu2016cluster, barroso2022cluster} or to programs that implement the cluster expansion method \cite{van2009multicomponent, seko2009cluster, troppenz2017predicting, chang2019clease, aangqvist2019icet, puchala2023casm}.

The cluster expansion method takes advantage of the fact that the potential site positions of atoms within a crystal are fixed, allowing for the energy of the crystal to be described in a format that enables rapid calculation. For instance, when considering a crystal structure composed of $N$ atoms from a material containing $M$ elemental species, the atomic positions in the crystal would typically be represented using $N$ real-valued 3-dimensional vectors. However, site locations can be expressed using an $N$-dimensional vector of $M$-valued discrete variables (referred to as the configuration vector $\vec{\sigma}$). In other words, within the cluster expansion method, energy $E$ is represented by a vector of discrete variables $\sigma_i$ (which for example takes values $0, 1, 2, ..., M - 1$), such that $E(\sigma_0, \sigma_1,...,\sigma_{N-1}) = E(\vec{\sigma})$. 
More specifically, the energy is expressed as an expansion of basis functions $\varphi_\alpha(\vec{\sigma})$ in the following form:
\begin{eqnarray}
E(\vec{\sigma}) = \sum_\alpha V_\alpha \varphi_\alpha(\vec{\sigma}), \label{eq:E_CE}
\end{eqnarray}
where $\varphi_\alpha(\vec{\sigma})$ is a polynomial composed of $\sigma_i$. Here, $\alpha$ represents the identifier of the basis functions, and $V_\alpha$ denotes the expansion coefficients. The $V_\alpha$ values are derived by fitting to the energies obtained from first-principles calculations, and various research groups have made program codes for this purpose publicly available \cite{van2009multicomponent, seko2009cluster, troppenz2017predicting, chang2019clease, aangqvist2019icet, puchala2023casm}. The key point of the cluster expansion method is that the $V_\alpha$ values learned from relatively small crystal structures are transferable to significantly larger crystal structures compared to the training data.

By transforming Eq.~\eqref{eq:E_CE}, it can be seen that $E(\vec{\sigma})$ is expressed as a polynomial of $\sigma_i$. That is,
\begin{eqnarray}
& &E(\vec{\sigma}) \nonumber \\
& & = A + \sum_i B_i \sigma_i + \sum_{i,j} C_{ij} \sigma_i \sigma_j + \sum_{i,j,k} D_{ijk} \sigma_i \sigma_j \sigma_k + \cdots,
\label{eq:E_CE_2}
\nonumber \\
\end{eqnarray}
In this paper, we consider the simplest case, a binary crystal structure, approximated up to the second term in the cluster expansion. In other words, for a crystal structure consisting of $N$ atoms, we have
\begin{eqnarray}
E(\vec{\sigma}) = A + \sum_{i=1}^N B_i \sigma_i + \sum_{i=1}^N \sum_{j>i} C_{ij} \sigma_i \sigma_j \label{eq:ce_2nd}
\end{eqnarray}
and consider the case with the binary values $\sigma_i = \pm 1$. This is the energy function of the so-called Ising model, and it is expected that optimization can be performed efficiently using Ising machines. In the next section, we will perform calculations for specific materials, the Au-Cu alloy, and verify their effectiveness.

Here, we summarize the computational methods and conditions used. The cluster expansion calculations for determining $V_\alpha$ in Eq.~\eqref{eq:E_CE} were performed using the Alloy Theoretic Automated Toolkit (ATAT) \cite{van2009multicomponent}, and the first-principles DFT calculations were carried out using the Vienna Ab Initio Simulation Package (VASP) \cite{kresse1993ab,kresse1994ab,kresse1996efficiency,kresse1996efficient,kresse1999ultrasoft}. In the VASP calculations, the Projector Augmented Wave method with the Perdew-Burke-Ernzerhof functional \cite{perdew1996generalized} was employed. The cutoff energy was set to 355 eV, and the $k$-point mesh was chosen with a 0.02\,\AA$^{-1}$ interval, including the $\Gamma$ point. The lattice constants and atomic positions were relaxed, with the convergence criterion set to a force of 0.02\,eV\AA$^{-1}$. The base cell for the cluster expansion method was a face-centered cubic (fcc) lattice with a side length of 3.8\,{\AA} and containing four atoms. The cluster dimension was limited to the second order, with a size constraint of 7.6\,{\AA} or smaller. Under these conditions, the maximum size of the training data for the cluster expansion method was a supercell containing 14 atoms.

\section{Results and Discussion}   \label{sec:results}

\begin{figure}[ht]
\centering
\includegraphics[width=\columnwidth]{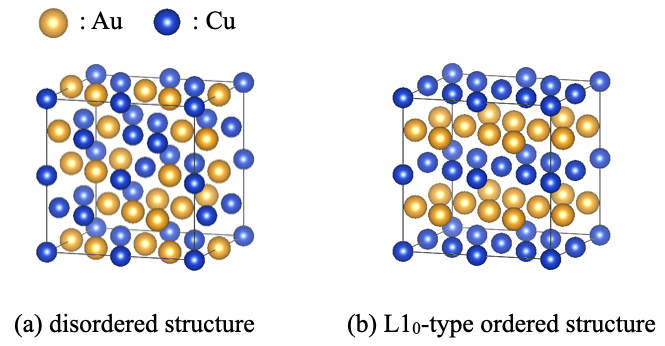}
\caption{The (a) disordered phase and (b) L1$_0$-type ordered phase of the Au-Cu alloy with a 1:1 composition ratio of Au and Cu, represented by a 32-site structure.}
\label{fig:AuCu}
\end{figure}

In this paper, we focus on the Au-Cu alloy as an example of a binary crystal structure. Detailed phase diagrams for this alloy system are well known \cite{fedorov2016cu}, and there are several examples of calculations using the cluster expansion method in the literature \cite{van2002automating, wei1987first, ozolicnvs1998cu, chang2019clease}. Both Au and Cu, as individual metals, adopt a fcc structure, and alloys composed of them form solid solutions at any Au-Cu composition ratio at high temperatures. The structure remains fcc and forms a disordered phase where Au and Cu randomly occupy each site. As the temperature decreases, several types of ordered phases appear, depending on the composition ratio \cite{fedorov2016cu}. As an example, Figure~\ref{fig:AuCu} shows the disordered phase and the ordered phase (below \SI{385}{\degreeCelsius}) for a 1:1 composition ratio. This ordered phase has a structure where layers of Au and Cu are stacked alternately. Note that VESTA\cite{momma2011vesta} is used for the visualization of crystal structures, including this figure, in this paper.

In order to derive the stable atomic arrangements for each composition ratio at low temperatures, it is necessary to find the optimal arrangement of Au and Cu atoms that yields the lowest energy for each composition. As a matter of fact, the case with a 1:1 composition ratio has the lowest energy \cite{ozolicnvs1998cu, chang2019clease}. This means that, when solving the optimization problem in Eq.\eqref{eq:ce_2nd}, the number of $+1$ and $-1$ values for $\sigma_i$ should be equal, and a solution corresponding to the stacked structure shown in Fig.\ref{fig:AuCu}(b) should be obtained. This is a so-called QUBO problem, and it can be solved directly using Ising machines. We will first investigate whether this simplest problem setting can be efficiently and successfully solved.

\begin{figure}[ht]
\centering
\includegraphics[width=\columnwidth]{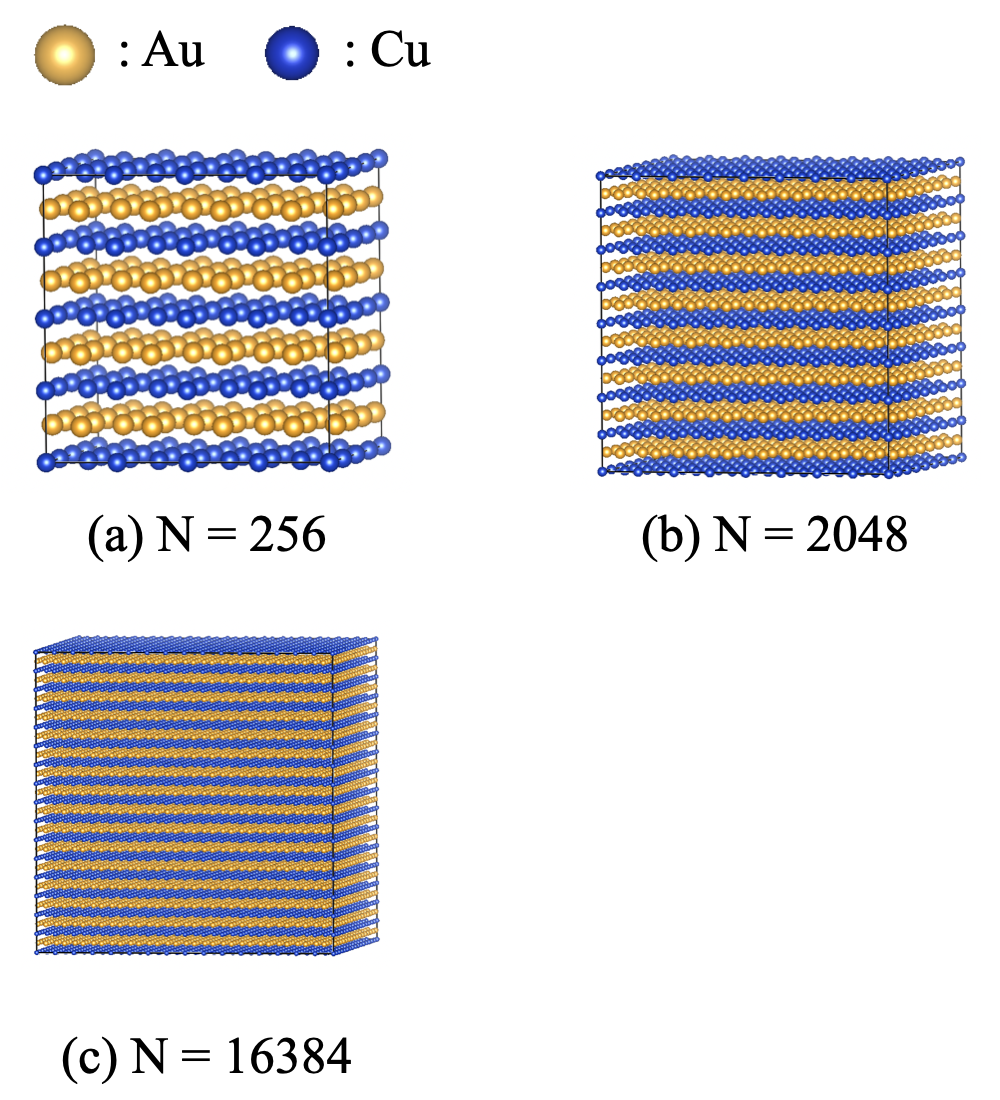}
\caption{The solutions for unconstrained optimization of the Au-Cu alloy with the number of atoms $N$: (a) 256, (b) 2048, and (c) 16384.}
\label{fig:AuCu_largeN}
\end{figure}

The expansion coefficients for Eq.~\eqref{eq:E_CE} obtained in the way as described at the end of the previous section were converted into the coefficients $A, B, C$ of Eq.~\eqref{eq:ce_2nd}, and optimization calculations were performed using the third-generation Fujitsu's Digital Annealer (DA) \cite{aramon2019physics}. The crystal structures targeted for optimization were cubic crystals obtained by multiplying each side of the base fcc lattice by 2, 4, 8, and 16, with each structure having $N = 32, 256, 2048, 16384$ atoms. The maximum number of atoms considered, 16384, is due to the third-generation DA's maximum handling capacity of approximately 100,000 bits (as shown in Eq.\eqref{eq:ce_2nd}, 1 bit per atom is required when representing the energy function of a binary system using a second-order approximation of the cluster expansion method). As a result of the optimization, the same optimal value of $E = -49.0$\,meV/atom was obtained for all structures, and the optimal solutions corresponded to crystal structures with an equal number of $\sigma_i = +1$ and $-1$. 
For the case of $N=32$, the solution is the layered structure shown in Fig.~\ref{fig:AuCu}(b), and the solutions for the cases of $N = 256, 2048, 16384$ are also same layered structures. For reference, these are illustrated in Fig.~\ref{fig:AuCu_largeN}.
This is in line with the expectation mentioned earlier and demonstrates that a valid solution was obtained using the Ising machine. 
In all cases, optimization was performed from several random initial solutions, and the same optimal value and solution were obtained.

Regarding the computation time required for optimization using DA, it took approximately 0.5 seconds for $N=32$ and 256, 1 second for $N=2048$, and $20 \sim 200$ seconds for $N=16384$. To compare computation time, a mathematical optimization solver (Gurobi Optimizer ver. 9.12 \cite{gurobi}) was used to obtain the solution. The time required to reach the same optimal value as DA was similar for $N=32$ and 256, taking less than 1 second, but for $N=2048$, it took $50 \sim 270$ minutes, suggesting that DA can efficiently solve large-scale problems.

Next, we consider optimization calculations under a given composition ratio of Au and Cu, which is of more interest in material science computations. This corresponds to imposing a constraint on the solution so that a fixed number of $\sigma_i$ become $+1$. Specifically, if we set $\sigma_i = +1$ when there is Au at site $i$ and $\sigma_i = -1$ when there is Cu, then for a solution with the composition ratio ${\rm Au}:{\rm Cu} = r:1-r$, the constraint condition is $\sum_i \sigma_i /N = r$. In this paper, we seek a solution that satisfies the constraint by optimizing Eq.~\eqref{eq:ce_2nd} with the constraint term $k \cdot \left( \sum_i \sigma_i -r N\right)^2$ added. Here, $k$ is a constant that determines the magnitude of the constraint term, and some adjustment is necessary to obtain a solution that satisfies the constraint.

\begin{figure}[ht]
\centering
\includegraphics[width=\columnwidth]{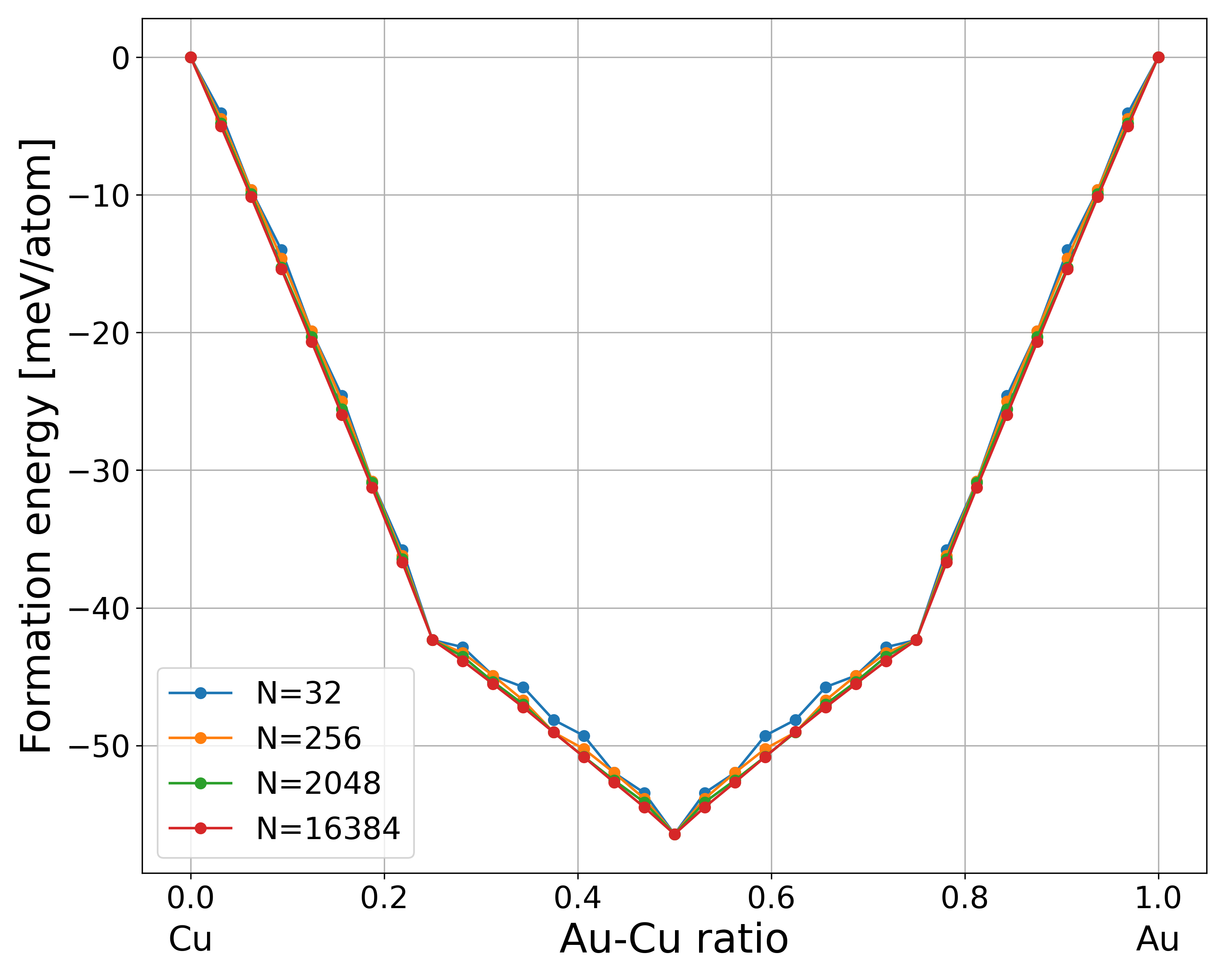}
\caption{Formation energy for each composition ratio when optimization calculations are performed under the constraint of composition ratio ${\rm Au}:{\rm Cu} = r:1-r$.}
\label{fig:occ_energy}
\end{figure}

\begin{figure}[ht]
\centering
\includegraphics[width=\columnwidth]{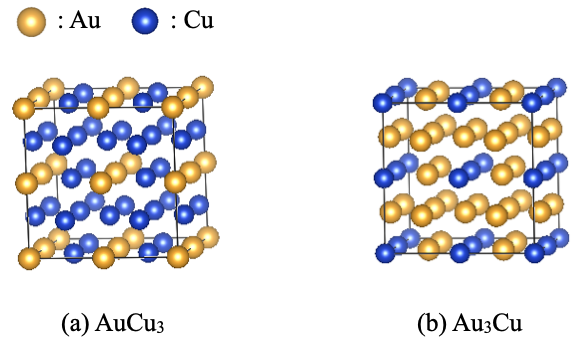}
\caption{For the Au-Cu alloy, the figure shows the ordered phases in a 32-site structure: (a) with an Au ratio of 0.25, and (b) with an Au ratio of 0.75.}
\label{fig:AuCu3_Au3Cu}
\end{figure}

The results of the constrained energy function optimization calculation are shown in Fig.~\ref{fig:occ_energy}. Here, instead of showing the total energy $E$ itself, we present the formation energy $\Delta E$ for the reaction \ce{$r$Au + $(1-r)$Cu -> Au_{r} Cu_{1-r}}. The conversion formula is $\Delta E(r) = E(r) - r E(r=1) - (1-r) E(r=0)$, where $E(r=1)$ and $E(r=0)$ correspond to the total energies of pure Au and Cu, respectively.
 The calculations were performed for the structure with $N = 32, 256, 2048, 16384$, and almost the same results were obtained. First, for $r=0.5$, the same energy value (converting to the formation energy, $\Delta E = -56.4$\,meV/atom) and solution as in the unconstrained case were obtained. Then, for $r=0.25$ and 0.75, in all cases with different numbers of atoms, we obtain $\Delta E = -42.3$\,meV/atom.
The obtained solutions for both cases are the L1$_2$-type ordered phases, as shown in Figure~\ref{fig:AuCu3_Au3Cu}(a) and (b), which is consistent with the literature~\cite{fedorov2016cu}. Examining the entire formation energy plot, the line segment connecting the points at $r=0, 0.25, 0.5, 0.75, 1$ forms the so-called convex hull. In other words, the formation energy at points other than these is slightly higher than the convex hull, by an order of $O(0.1) \sim O(1)$ meV/atom. Since the points on the convex hull represent thermodynamically stable states, this result indicates that only the compositions at $r=0, 0.25, 0.5, 0.75, 1$ form stable phases in the ground state. This result is consistent with the known phase diagram~\cite{fedorov2016cu}, demonstrating the successful optimization by DA.

As mentioned earlier, nearly the same results were obtained for different values of $N$. However, upon closer examination, it is observed that for points other than $r=0, 0.25, 0.5, 0.75, 1$, the energy value decreases as $N$ increases. This is because, at those points, several phases coexist (for example, at $r=0.375$, \ce{AuCu3} and \ce{AuCu} coexist), but the correct phase separation pattern cannot be represented when $N$ is small. Conversely, the ability to perform optimization for the larger $N$ cases demonstrated in this study allows us to obtain more accurate insights into the separation patterns of mixed phases at various composition ratios, which is highly valuable.

Lastly, we discuss the calculation time for the case with constraints. In the case with constraints, a noticeable difference was observed between DA and Gurobi, even for $N=256$. Comparing the average calculation times for all composition ratios, DA took 0.51 seconds, while Gurobi took 2.3 seconds. Furthermore, in some cases (4.4\% of the total), Gurobi's optimization calculation did not reach the optimal value obtained by DA even after 60 seconds. For $N=2048$, the average calculation time for DA was 3.9 seconds, while Gurobi's was 3300 seconds, and in 41.2\% of the cases, Gurobi's optimization calculation did not reach the value of the DA's solution even after 360 minutes. Note that for $N=16384$, the average solving time for DA was 241 seconds (Gurobi calculations were not performed because they were expected to take a very long time).

\section{Conclusion}  \label{sec:conclusion}
In this paper, we demonstrated that the optimal elemental configuration search problem, which involves determining which sites on a given lattice are energetically stable for each atomic species in an Au-Cu alloy, can be efficiently solved using an Ising machine by converting the energy function into a QUBO function using the cluster expansion method. In particular, plausible solutions were obtained even for optimization problems with constraints on the solutions, such as when the composition ratio of atomic species is given, indicating that this approach is useful for problems of interest in materials science.

Furthermore, thanks to the recent significant improvements in the performance of Ising machines due to advancements in both hardware and software, it has been demonstrated that calculations with a large number of atoms can be performed in a short amount of time. In the third generation of Fujitsu Digital Annealer, about 100,000 fully connected bits are available, and in this paper, optimization was performed for up to 16384 atoms. Even when compositional constraints were imposed, the optimization was completed in just a few minutes at most, and plausible solutions corresponding to crystal structures known from experimentally obtained phase diagrams were obtained. Although this result was expected, it is still remarkable. This largest-scale structure corresponds to a cube with an edge length of approximately 6\,nm, representing a nanoscale model. The ability to perform atomistic simulations of such large-scale models is a first step in connecting atomic-scale phenomena with macroscale phenomena, and it is expected to serve as a new tool for future materials research.

In this study, we dealt with the simplest case of approximating the energy function of a binary crystal using the second-order cluster expansion method. For this approach to become more useful in the future, various technical advancements are desired. First, it is common in cluster expansion methods to include terms up to the third or fourth order, which, as in Eq.\eqref{eq:E_CE_2}, introduces terms of the third or higher order for binary variables. To make these terms computable with an Ising machine, they need to be transformed into terms of the second order or lower, which usually requires the introduction of auxiliary bits and constraint terms. This results in the use of a large number of bits compared to the number of atoms and increased difficulty in solving the problem. Therefore, more efficient handling of higher-order terms is required to deal with large atomic models while improving the accuracy of cluster expansion approximations. Another challenge is addressing crystal structures of ternary systems or higher (elemental configuration optimization problem with three or more atomic species). In conventional cluster expansion methods, this is formulated using $M$-value discrete variables ($M \ge 3$), which requires assigning multiple bits to a single site for Ising machines, again necessitating the use of a large number of bits compared to the number of atoms. In Ref.\cite{choubisa2023accelerated}, an extended method for the conventional cluster expansion has been proposed to handle multi-component systems with Ising machines, solving the search problem for up to eight atomic species using DA. Such new problem formulations, bit reduction techniques, and further large-scale and high-speed Ising machines are expected to make Ising machines more useful tools in computational materials science.

\bibliography{references} 

\end{document}